\newcommand{\xdeleted}[1]{}

\documentclass[astronomy,article,accept,pdftex,moreauthors]{Definitions/mdpi}

\usepackage[normalem]{ulem}
\firstpage{1} 
\pubvolume{1}
\issuenum{1}
\articlenumber{0}
\pubyear{2025}
\copyrightyear{2025}
\externaleditor{Firstname Lastname} 
\datereceived{27 August 2025} 
\daterevised{30 November 2025} 
\dateaccepted{1 December 2025} 
\datepublished{ } 
\hreflink{https://doi.org/} 
\Title{Disentangling the Cosmic/Comoving Duality: The Cognitive Stability and Typicality Tests}

\TitleCitation{Disentangling the Cosmic/Comoving Duality: The Cognitive Stability and Typicality Tests}

\Author{Meir Shimon $^{1,2}$
 \orcidA{}}

\AuthorNames{Meir Shimon}

\AuthorCitation{Shimon, M.}

\address{%
$^{1}$ \quad School of Physics and Astronomy, Tel Aviv University,
Tel Aviv 69978, Israel; meirs@tauex.tau.ac.il\\
$^{2}$ \quad Afeka Tel-Aviv Academic College of Engineering, Tel Aviv 69107, Israel}

\abstract{Cosmological scenarios wherein the cumulative 
number of spontaneously formed, cognitively impaired, disembodied 
transient observers is vastly larger than the corresponding number of 
atypical `ordinary observers' (OOs) formed in the conventional way---essentially 
via cosmic evolution and gravitational instability---are disqualified in modern cosmology on the 
grounds of Cognitive Instability---the untrustworsiness of one own's reasoning---let alone the atypicality of OOs like us.
According to the concordance $\Lambda$CDM cosmological model---when described in the (expanding) `cosmic frame'---the cosmological expansion is future-eternal. In this frame we are atypical OOs, which are vastly outnumbered by typical Boltzmann Brains (BBs) that spontaneously form via sheer thermal fluctuations in the future-eternal asymptotic de Sitter spacetime.
In the case that dark energy (DE) ultimately decays,  the cumulative number of transient `Freak Observers' (FOs) formed and destroyed spontaneously by virtue of the quantum uncertainty principle ultimately overwhelms that of OOs. Either possibility is unacceptable. 
We argue that these unsettling conclusions are artifacts of employing the (default) 
cosmic frame description in which space expands.  When analyzed in the comoving frame, OOs overwhelmingly outnumber both BBs and FOs. 
This suggests that the dual comoving description is the cognitively stable preferred framework for describing our evolving Universe. 
In this frame, space is globally static, masses monotonically increase, and the space 
describing gravitationally bounded objects monotonically contracts.} 


\begin{document}

\section{Introduction}\label{sec1} 

The concordance flat $\Lambda$CDM cosmological model successfully 
and efficiently describes our Universe from very early 
after the Big Bang until the present time, and from the Hubble 
scale down to galaxy cluster scales, and it does so with 
only a handful of free parameters. Yet, it has a certain disturbing 
feature that has long been recognized and acknowledged.

According to the concordance $\Lambda$CDM model---which contains cold dark matter (CDM) and a cosmological 
constant, $\Lambda$---the current
accelerated expansion phase of the observable Universe will 
last indefinitely, and we appear to be very `special,' i.e., atypical,
observers, since we make our observations at a finite time
after the beginning of a future-eternal Universe.
This somewhat counter-intuitive conclusion arises naturally
within the concordance model and is closely related to another
well-known puzzle---the `why now?' problem. The latter revolves 
around the question of how likely is it for us to observe the 
Universe in the unique era in which dark energy (DE) and 
non-relativistic (NR) matter contribute comparably to the 
cosmic energy budget, in spite of their very different evolution 
histories. This is the Cosmic Coincidence Problem (CCP) 
\cite{Steinhardt1997,Carroll2001,Weinberg2008}.

Whereas the common expectation that we are typical observers has been challenged~\cite{HartleSrednicki2007}, 
compelling counter-arguments in favor of our 
typicality, namely that our observations are typical within our reference class, e.g., 
\cite{Gott2008,Page2007},
touch upon the bedrock of the scientific method:
when experimentally or observationally tested theory is compared to only a finite number of data points. 
It is a fundamental underlying assumption that is made by default that 
these data points (i.e., observations) are 'typical` and randomly selected 
in the observationally accessible range. Otherwise, we cannot trust our conclusions that 
theory provides a good fit to the data, for there is always a non-negligible chance in 
this case that the measurements are carried out specifically where theory matches the data, 
and fails otherwise.

In a future-eternal Universe, any process with a finite (however small, yet finite) 
probability to take place will do so an infinite number of times.
In $O(10^{110})$ years, after the most massive black holes have been 
evaporated, and long after the last luminous stars have exhausted 
their nuclear fuel, the Universe is expected to enter a Dark Era 
 described by a static de Sitter spacetime containing 
a heat bath of massless photons, gravitons, and other light 
degrees of freedom, characterized by a Gibbons--Hawking (GH) temperature 
$K_{B}T_{GH}=\frac{\hbar c}{2\pi}\sqrt{\frac{\Lambda}{3}}$ 
\cite{GibbonsHawking1977}, where $K_{B}$, $\hbar$ and $c$ are the Boltzmann 
constant, the reduced Planck constant, and speed of light, respectively. 
Occasionally, and due to 
the infinitely long time available for this (otherwise unlikely) 
event to take place, very rare thermal fluctuations
will randomly form a certain type of transient `Boltzmann Brains' (BBs), 
for a very brief time period, but with highly disordered memories and
cognitive disabilities, e.g., \cite{AlbrechtSorbo2004,Carroll2017}.
The latter refers to the unreliability of one own's 
reasoning and judgement abilities---basically, theories that predict their own demise and untrustworthiness.
These transient BB observers might even think that they live 14 Gyrs 
after the Big Bang in an ordered Universe, where large-scale structures 
are formed via gravitational instability in the conventional way, etc., 
but this thought would not last for longer than a brief moment. 
BBs---rather than Ordinary Observers (OOs) like ourselves who formed 
in the usual way---are expected to be typical observers in an eternal 
Universe such as the one which is described by the concordance 
model \cite{Page2007}. This is so because the cosmological constant 
comes to dominate the expansion a finite time after the beginning.
After this transition takes place, the Universe enters an 
infinitely long period of exponential expansion in which essentially 
no new structure forms that evolves---among other structures---to habitable~planets. 

Even if DE ultimately decays, eventually leaving behind an empty Milne-like 
spacetime, one expects yet another type of cognitively impaired 
`Freak Observers' (FOs) to form thanks to Heisenberg's 
Uncertainty Principle, e.g., \cite{Page2005}. FO---essentially an entire self-aware brain---can 
spontaneously pop into existence, with all memories built in,
giving it a false impression of `reality' as we know it.
This channel of FO production typically has a much larger
(albeit still stupendously small) production rate than BBs 
do. Yet, in a future-eternal Universe (either 
asymptotically $\Lambda$-dominated de Sitter or empty Milne), the number of 
untrustworthy observations made by either type---BBs or FOs---will overwhelm the number of observations made by OOs. Since 
only a finite narrow time window is available for the latter type to 
form, basically until DE takes over the expansion rate, it is then 
paradoxical that we seem to be belonging to the class of OOs rather 
than to either BBs or FOs.

This paradox has a long history starting as early as in the wake of the 
nineteenth century, with Boltzmann conception of an entire expanding 
Universe that starts off as a thermal 
fluctuation, through a latter idea by Eddington that it is even much more 
likely that a `mathematical physicist' forms via (a much smaller and 
therefore more probable) thermal fluctuation \cite{Eddington1931}, and 
the more modern way of thinking about it as the minimal randomly formed 
organism necessary for observing the world---a BB, e.g., 
\cite{AlbrechtSorbo2004,Carroll2017}. In spite of the significant 
progress achieved over the years in the field of cosmology, and multiple attempts 
to address the BB paradox, e.g., \cite{Page2005, Carlip2007,BoussoFreivogel2007,Gott2008, Page2009,Albrecht2009,DavenportOlum2010,BoddyCarroll2013,Boddy2015,Horvat2015,
Page2017}---in some cases along with its FO variant---it remains defiant and still largely regarded an open question. 
It was pointed out that this could simply be a measure problem in the multiverse context, e.g., 
\cite{Vilenkin2007,Bousso2008,Bousso2009,deSimone2010,Salem2012,
Olum2021}. However, 
the discussion in the present work focuses on the non-OO problem of our own Universe, 
that is independent of whether the multiverse exists or not.

The observation that BBs' and FOs' abundance overwhelms that of OOs 
challenges the assumption of typicality in cosmology, 
as it implies that most observations in an eternal Universe would 
be made by disembodied brains rather than by OOs 
like us. Even worse, these (standard model) scenarios are justifiably 
subject to criticism on the grounds of Cognitive Instability; scenarios 
in which the most likely observers have disordered false memories 
and incoherent thoughts cannot possibly be trusted and qualify as 
viable cosmological models, e.g., \cite{Winsberg2012,Carroll2017,Avni2022,Elga2025}.

This argument, and possibly
others as well, renders the presence of BBs and FOs 
a diagnostic tool, rather than a realistic possibility; 
a scenario or model that predicts an overwhelming abundance 
of either BBs or FOs over OOs is deemed unfeasible 
or pathological. This is currently debated in the context of the
prevailing inflationary scenario---eternal inflation---and the multiverse. The problem with
this Cognitive Instability test is that even our
own, $\Lambda$CDM Universe, is future-eternal and therefore
must be plagued by the BB (or FO) problem, even if we ignore inflation
and the multiverse \cite{Page2005}; this problem still afflicts our
single $\Lambda$CDM Universe.
Therefore, it seems---based on these arguments---that either
we are very atypical and rare OOs \cite{Page2008}
or the Universe must come to an end, i.e., it decays
\cite{Page2005}, or faces a `cosmic doomsday', e.g., \cite{Page2008,Page2008b}. 
None of these possibilities are very appealing, especially since 
they are proposed first and foremost to explain the (otherwise 
almost inconceivable) existence of OOs like us. 
In the present work a much more mundane explanation is proposed 
that does not by itself recourse to new physics.

This paper is organized as follows. 
In Section~\ref{sec2} the cosmic/comoving duality 
in the standard cosmological model is described. 
In Section~\ref{sec3} the conventional considerations in the cosmic frame that lead to the perplexing BB and FO paradoxes 
are laid out. In Section~\ref{sec4} it is shown how these paradoxes go away when analyzed in the comoving frame. This work is summarized in 
Section~\ref{sec5}, where it is concluded that, at least from the perspective of 
the BB and FO paradoxes, the cognitively stable picture---and one that does not run 
into typicality issues---is given by the comoving frame description.

\section{The Cosmic/Comoving Frame Duality in Cosmology}\label{sec2}

A homogeneous and isotropic spacetime is described in the cosmic (expanding) 
frame by the Friedmann--Robertson--Walker (FRW) infinitesimal line interval, 
$ds^{2}=g_{\mu\nu}dx^{\mu}dx^{\nu}$, where summation convention over 
the spacetime coordinates applies,
\begin{eqnarray}\label{eq:2.1}
ds^{2} &=& -dt^{2} + a^{2}(t)
\left[\frac{dr^{2}}{1-Kr^{2}} + r^{2}(d\theta^{2}
+ \sin^{2}\theta\, d\varphi^{2})\right] \nonumber \\
&=& a^{2}(\eta) \left[-d\eta^{2} + \frac{dr^{2}}{1-Kr^{2}} + r^{2}(d\theta^{2}
+ \sin^{2}\theta\, d\varphi^{2})\right],
\end{eqnarray}
and $t$ and $\eta$ are the cosmic and
conformal time coordinates, respectively, which 
are related via $d\eta \equiv \frac{dt}{a}$, $a(t)$ is the scale factor,
and $K$ is the spatial curvature parameter. By convention, $a_{0}$, the scale factor 
at present, is set to unity. The scale factor is related to the redshift observable via $a=\frac{1}{1+z}$. Space is expanding in this frame 
and particle masses are fixed universal constants. The scale factor $a(t)$ is determined 
by straightforward integration of the Friedmann equation:
\begin{eqnarray}
H=H_{0}\sqrt{\frac{\Omega_{r}}{a^{4}}+\frac{\Omega_{m}}{a^{3}}+\frac{\Omega_{k}}{a^{2}}+\Omega_{\Lambda}}, 
\end{eqnarray}
where $H\equiv d\ln(a)/dt$ is the expansion rate, $H_{0}$ is its current value, and $\Omega_{r}$, $\Omega_{m}$, 
$\Omega_{k}\equiv -Kc^{2}/H_{0}^{2}$ and $\Omega_{\Lambda}\equiv \Lambda c^{2}/(3H_{0}^{2})$ 
are the (cosmologically averaged) energy densities in critical density 
units of radiation (relativistic degrees of freedom such as the cosmic microwave background photons), NR matter (baryons and CDM), 
the effective energy density associated with the global spatial curvature parameterized by $K$, and the cosmological constant, respectively, subject to the constraint $\Omega_{r}+\Omega_{m}+\Omega_{k}+\Omega_{\Lambda}=1$. 
Here, we assume the energy content of either flat $\Lambda$CDM or K$\Lambda$CDM 
(extension of flat $\Lambda$CDM allowing for non-flat spatial geometries, $K\neq 0$).

Scaling the $a^{2}(\eta)$ factor out from the FRW metric, we
are left with a static metric $d\tilde{s}^{2}=\tilde{g}_{\mu\nu}dx^{\mu}dx^{\nu}$:
\begin{eqnarray}
d\tilde{s}^{2} &=& -dt^{2}/a^{2}(t) + \frac{dr^{2}}{1-Kr^{2}} + r^{2}(d\theta^{2}
+ \sin^{2}\theta\, d\varphi^{2}) \nonumber \\
&=& -d\eta^{2} + \frac{dr^{2}}{1-Kr^{2}} + r^{2}(d\theta^{2}
+ \sin^{2}\theta\, d\varphi^{2}),
\end{eqnarray}
which describes spacetime in the comoving frame, where it is understood that the time coordinate here is $\eta$. 
The time dependence of the scale factor in the cosmic frame is replaced by time-dependent masses when described in the comoving frame, \linebreak$m(\eta) = m_{0} a(\eta)$. This is the (indeed less popular) 
alternative comoving frame picture that is mostly employed as a computational 
tool, e.g., in~\cite{HwangNoh2001, LesgourguesPastor2006}, 
and in the Mukhanov--Sasaki equation of quantum perturbations of the inflaton field, 
as well as in standard Boltzmann solvers, e.g., CAMB. Establishing the formal 
duality---essentially indistinguishability---between these two alternative 
descriptions (at the background level as well as at the linear perturbation level) is straightforward, e.g., \cite{Deruelle2011, Bars2014, Lombriser2023, Shimon2025}. In both pictures, space relatively expands; 
in the cosmic frame, it expands relative to fixed yardsticks---the Planck length and 
Compton wavelength of, e.g., the electron. In the comoving frame, space is static but the yardsticks contract.

In the standard model (SM) of particle physics, masses are fixed universal 
constants by convention. This is manifested by the fact that 
the Higgs vacuum expectation value (VEV) and the chirality breaking 
scale in quantum chromodynamics (QCD) are fixed constants. 
The universal constant of gravitation, $G$, and consequently 
the Planck mass $m_{pl}=G^{-\frac{1}{2}}$, is fixed as well. 
Again, this is no more than a 'convenient' convention; 
if all these masses would vary identically in space and time, 
there is practically no experimental or observational way in which 
the two alternative pictures could be distinguished.

Allowing for masses to vary, in particular multiplying them by 
a scale function $m_{0}\rightarrow m_{0}a(t)$ in the presence of the spacetime metric 
(that is necessarily present in the SM lagrangian), implies that the corresponding 
VEV of the Higgs field must be locally rescaled to offset that of the metric; 
masses now carry the time dependence that would be otherwise carried by 
the expanding spacetime metric. This implies that the SM of particle physics must be `conformalized' ~\cite{Mannheim2016}, otherwise no compensation can be introduced to counteract the new contributions from the time derivatives of the Higgs VEV. The QCD sector is similarly conformalized. This framework was established in~\cite{Bars2014}. 
It certainly represents a radical departure from the standard convention of constant VEVs, but similar concepts have been explored in the literature, e.g.,~\cite{Uzan2003,Wetterich2013,Lombriser2023,Shimon2025,Mannheim2016}, and indeed this is no more and no less than an issue of convention; carefully conformalizing the SM of particle physics and general relativity (GR), all masses have the same spacetime evolution and so measurable dimensionless mass ratios are identical to those measured based on the standard fixed mass~convention.

A point of conceptual (albeit no practical) importance is that in the process of conformalization of the SM of particle physics, the Higgs sector Lagrangian density is corrected by a conformal coupling term of the curvature scalar $\frac{1}{6}RH^{\dagger}H$ where $R$ and $H$ are the Ricci curvature scalar and $H$ is the Higgs field ~\cite{Bars2014}. The purpose of adding this term is to balance spacetime variations of the Higgs VEV in its kinetic term so that overall the action is locally scale-invariant. Comparing this new term to the effective mass term in the Higgs sector, $m_{H}^{2}H^{\dagger}H$, where $m_{H}$ is the Higgs mass, and recalling that $R\sim G\rho$, we conclude that the new curvature term essentially does not affect the Higgs mass value because its fractional contribution, even in as dense regions as a typical neutron star density, $\rho\lesssim m_{p}^{4}$, where $m_{p}$ is the proton mass, is 
$G\rho/m_{H}^{2}\lesssim m_{p}^{4}/(m_{pl}m_{H})^{2}\lesssim 10^{-40}$.  

Going beyond the homogeneous and isotropic approximation, i.e., adding linear scalar metric perturbations in the FRW line element, Equation~(1) generalizes to
\begin{eqnarray}
ds^{2}=a^{2}(\eta)[-(1+2\Phi)d\eta^{2}+(1-2\Phi)\gamma_{ij}dx^{i}dx^{j}], 
\end{eqnarray}
where $\Phi$ is the gravitational potential and $\gamma_{ij}$
is the spatial metric. However, in the cosmic frame description, light emitted at
a given time will undergo redshift irrespective of whether the source
is gravitationally bound or is at rest at the Hubble frame. 
Therefore, it is essential to show how light emitted from 
gravitationally bound objects in the comoving frame description is still redshifted 
even though space does not expand from this perspective.

Considering nonlinear gravitationally bounded objects, if the spatial coordinates are transformed 
$dx^{i} \rightarrow d x^{i} / a$ (where the superscript $i$ runs through 
all three spatial coordinates), while masses scale as 
$m=m_{0}a$ as in the background of the comoving frame, 
then emitted light from nonlinear gravitationally bound objects is redshifted, irrespective of whether the emitting atoms, molecules, etc., reside in the background 
or in gravitationally bounded objects. This is so because the descriptions of 
both the background and of that of gravitationally bound objects 
is conformally transformed in going from the cosmic to comoving frame, and null geodesics, subject to the constraint $ds^{2}=0$, are blind to conformal 
transformations, e.g.,~\cite{Padmanabhan2010}. To summarize, in the comoving frame, a nonlinear gravitationally bound object, e.g., a galaxy, star, planet, etc., is described in the comoving frame by a metric of the form
\begin{eqnarray}
d\tilde{s}^{2}=a^{-2}(t)[-dt^{2}+dr^{2} + r^{2}(d\theta^{2}
+ \sin^{2}\theta\, d\varphi^{2})],
\end{eqnarray}
where we ignored additional spatio-temporal metric variations via evolving gravitational potentials but rather imprinted 
the effect of the global cosmic evolution on the spacetime metric that describes 
these structures (that would be considered static in the cosmic frame picture).
The nontrivial lapse function $g_{tt}=a^{-2}$ that is common to both the background spacetime (Equation~(3)) and that of gravitationally bound objects (Equation~(5)) is responsible for the gravitational redshift according to the classical general--relativistic redshifting of frequencies 
$\nu\propto\sqrt{g_{tt}}=a^{-1}=1+z$.
Thus, the redshift phenomenon that is naturally explained in the cosmic frame is
equally well explained in the comoving frame description,
but now gravitationally bounded objects actually contract over time relative 
to the static background, and the static background Universe appears to be expanding
only relative to these contracting objects. This includes our
own solar system, the Sun, Earth, etc., from within which 
we observe the Universe to be relatively expanding. 
In any case, both cosmic and comoving frame descriptions are 
conformally related and null geodesics, subject to the constraint 
$ds^{2}=0$, and are unaffected when switching between the cosmic and comoving frame descriptions. 
This fact is significant as our cosmological observations are limited to 
the past lightcone.

This seemingly innocuous difference between the standard (i.e., cosmic) and the comoving descriptions
makes a crucial difference for the FO and BB problems, as we see below. 
The key reason for this difference is the disparity between the 
roles played by masses and volumes
in the production rate of FOs and BBs; whereas the volume appears as
an overall multiplicative factor (the number of observers plausibly 
scales linearly with the volume), the mass appears in a Boltzmann suppression factor in the expression for the production rate, i.e., the likelihood for production decays exponentially with the mass. In both thermal and quantum scenarios, the respective BBs and FOs are overwhelmingly produced 
during the indefinite phases of expansion in the cosmic frame description.
In contrast, in the comoving frame description, the corresponding
production rates are characterized by growing brain
masses, which exponentially outweighs the contraction of 
brains (considered bound objects in our analysis). 
This will be discussed in~Section~\ref{sec4}.

\section{The Standard (Cosmic Frame) Argument and the Atypicality of OOs}\label{sec3}

In the following, we closely follow the cosmic frame-based derivation of ~\cite{Page2005}. The core idea is that no matter how unlikely the event of the spontaneous formation of transient disembodied brains to occur is, insofar the probability is finite it will take place an infinite number of times in an indefinitely expanding Universe. In contrast, the conventional formation of OOs essentially halts once DE takes over the expansion, a finite time after the Big Bang. This 'nightmare scenario' leads to the BB (and FO) paradox.

Since an FRW spacetime has a horizon it then follows that it has a Hawking temperature, $T_{H}$. Specifically, e.g.,~\cite{Padmanabhan2003,CaiKim2005,AkbarCai2007,Cai2009},
\begin{eqnarray}
T_{H}=\frac{\hbar c}{2\pi k_{B}}\sqrt{H^{2}+\frac{K}{a^{2}}}. 
\end{eqnarray}
Considering $\Lambda$CDM, or even the K$\Lambda$CDM model, a 
sufficiently long time after the various contributions 
(including that of spatial curvature) redshifted away 
$H^{2}\rightarrow \Lambda/3$ and we arrive at the Gibbons--Hawking (GH) 
temperature of de Sitter spacetime, $K_{B}T_{GH}=\frac{\hbar c}{2\pi}\sqrt{\frac{\Lambda}{3}}$ \cite{GibbonsHawking1977}.

The cumulative number of BBs of mass $m_{0}$ to form is essentially
proportional to the product of the Boltzmann factor, 
$e^{-\alpha_{BB}}$, where 
$\alpha_{BB}\equiv m_{0}c^{2}/(k_{B}T_{GH})=2\pi m_{0}c/(\hbar\sqrt{\Lambda/3})$, and the time-integrated three-dimensional spatial volume of the observable Universe in units 
of the spacetime brain volume. Other, non-exponential factors, are ignored
because the overwhelmingly dominant term is the Boltzmann factor,
due to the extraordinarily huge $\alpha_{BB}$.
The cumulative number of such brains throughout the infinite lifetime of the Universe is
\begin{eqnarray}
N_{BB}\sim\int_{t_{0}}^{\infty}\frac{\exp(-\alpha_{BB})a^{3}(t) dt}{V_{br}\tau},
\end{eqnarray}
where $t_{0}$ is the present time when structure formation essentially 
ceases and consequently new OOs stop forming, and $V_{br}$ is the three-dimensional brain volume. 
Assuming flat space, the volume is infinite and so the number of BBs and OOs will be infinite 
as well. To regularize these infinities it is customary to assume a finite comoving 
volume integrated indefinitely over the eternal lifetime of the $\Lambda$CDM Universe.
Since the scale factor 
$a(t)\sim e^{\sqrt{Ht}}$ grows exponentially during 
the $\Lambda$-domination era with $H=\sqrt{\frac{\Lambda}{3}}$, then $dt=da/a$ 
at late times and so 
$N_{BB}\sim\int_{1}^{\infty}\frac{\exp(-\alpha_{BB})a^{2} da}{V_{br}\tau}=
\frac{\exp(-\alpha_{BB})}{V_{br}\tau}\int_{1}^{\infty}a^{2} da$ 
diverges, however large (yet finite) $\alpha_{BB}$ is. Here, the lower integration limit is set at $a=1$ for concreteness, although in practice, this limit will be higher. In any case, the divergence comes from the upper integration limit, so we ignore this subtlety in what follows. 
Crucially for the current discussion, and in contrast to the corresponding calculation in the comoving frame 
discussed in the next section, the extremely small (but finite) Boltzmann term decouples from 
the diverging term $\int_{1}^{\infty}a^{2} da$, thereby giving rise to diverging $N_{BB}$.
This results in the absurd expectation for the thermal fluctuation into and out of existence of an infinite number of BBs over the future-eternal lifespan of the Universe. 
Estimates of the human population to have ever lived on Earth are in the ballpark of $10^{11}$ humans. Even if we make the bold and entirely unrealistic assumption that 
the number of habitable planets that eventually 
formed self-aware life is as large as the number of luminous stars in the 
observable Universe, $O(10^{22})$, we still end up 
with $O(10^{33})$ OOs at best (assuming that the cumulative Earthly population is representative among all habitable planets). This wildly generous upper limit on 
the number of OOs ever formed via conventional cosmic evolution is still 
finite and therefore vastly swamped by the diverging $N_{BB}$.   

In the case that $\Lambda$CDM is only an over-idealized approximation, and DE is not accounted for by a pure gravitational constant, and that it ultimately decays, then after a sufficiently long (but finite) time 
the Universe approaches an empty spacetime state, essentially 
a Milne Universe with $a=\sqrt{-K}t$ that is characterized by spatially open geometry, $K<0$.
In this case, $dt=da/\sqrt{-K}$, and the Hawking temperature vanishes, $T_{H}=0$, following Equation~(6). 
Although this implies that the BB formation channel does not exist in this case, transient deluded brains can still spontaneously form via the Heisenberg Uncertainty Principle. Following a similar rationale to the one that led to the estimated 
$N_{BB}$ in Equation~(7), the Boltzmann factor $\exp(-m_{0}c^{2}/T_{GH})$
is now replaced by $\exp(-m_{0}c^{2}\tau/\hbar)$, where $c$ is the speed of light, and $\tau$ is the brief lifetime 
of the brain that is assumed to be sufficiently long as 
to allow for a 'conscious observation' of the Universe. 

In this case, we obtain that 
\begin{eqnarray}
N_{FO}\sim\int_{t_{0}}^{\infty}\frac{\exp(-\alpha_{FO})a^{3}(t) dt}{V_{br}\tau}=\int_{1}^{\infty}\frac{\exp(-\alpha_{FO})a^{3}da}{\sqrt{-K}V_{br}\tau},
\end{eqnarray}
i.e., $N_{FO}\sim\frac{\exp(-\alpha_{FO})}{\sqrt{-K}V_{br}\tau}\int_{1}^{\infty}a^{3} da$, 
diverges, similar to the BB case, irrespective of how large $\alpha_{FO}$ is. 
As in the BBs' case, the lower limit of the integration we employ, $a=1$, is unrealistic---the present Universe is far from being empty. 
A more plausible lower integration limit is some $a\gg 1$ that corresponds to a scenario in which DE decays and the Universe empties out in the remote future. Nevertheless, exactly as in the BBs' case, the divergence comes from the upper limit on $a$ so our conclusions are unchanged---the cumulative number of FOs 
integrated along the indefinite expansion diverges and infinitely exceeds that 
of OOs.

To obtain a sense of how large $\alpha_{BB}$ and $\alpha_{FO}$ are, we make the following assumptions.
The mass of the disembodied brain is $m_{0}\sim$ 1 Kg, the contribution of DE (assuming that it is indeed accounted for by a cosmological constant) to the present-day cosmic energy budget 
in critical density units $\Omega_{\Lambda}=\Lambda/(3H_{0}^{2})=0.69$ ~\cite{Planck2018}, the current 
expansion rate is \mbox{$H_{0}=67$ km/sec/Mpc}, and the lifetime of the brain is $\tau=0.1$ s.
Under these assumptions we obtain
\begin{eqnarray}
\alpha_{BB}&=&\frac{2\pi M c}{\hbar H_{0}\sqrt{\Omega_{\Lambda}}}\sim 4.7\times 10^{68}\\
\alpha_{FO}&=&\frac{M c^{2}\tau}{\hbar}\sim 1.4\times 10^{49}.
\end{eqnarray}
These specific values were irrelevant to the conclusions arrived at in this section, nor will they be relevant to the next section's analysis. What will be important and indeed relevant is that $\alpha_{BB},\ \alpha_{FO}\gg 1$.

It is worth mentioning at this point that carbon-soft-tissue-based brains are not the simplest imaginable 
observers. {\it A priori}, future self-aware silicon-based human-made observers are not theoretically ruled out, 
and it is not entirely impossible that such artificial-like systems will spontaneously occur in the future de Sitter vacuum or 
empty Milne Universe. These are significantly less complex than a human brain and the probability for their spontaneous 
creation is much larger, e.g.,~\cite{Banks2007,Bousso2012}. Yet, as discussed in Section~\ref{sec4} below, when considered 
in the comoving frame, even this type of minimal disembodied conscious observer is overwhelmed by OOs.

\section{Typicality Restoration: The Comoving Frame Perspective}\label{sec4}

The comoving picture is obtained from the cosmic picture practically by setting $a=1$ in the metric and simultaneously applying the transformation 
$m_{0}\rightarrow m_{0}a(\eta)$. While the masses (including the Planck mass) 
change in this picture, the Hawking temperature solely depends on the geometry and is 
not affected by the new dynamics. 
This implies that in the comoving frame described by the metric of Equation~(3)
\begin{eqnarray}
\tilde{T}_{H}=\frac{\hbar c\sqrt{K}}{2\pi k_{B}}, 
\end{eqnarray}
and unlike in the cosmic frame, Equation~(6), it does not 
evolve. Following the standard assumption that $K=0$, we 
see that the Hawking temperature vanishes, whereas in the 
case $K<0$, it is not even defined; in both cases there is no horizon, 
and consequently no heat bath for BBs to form. Only in the case where 
$K>0$ does cosmic horizon exists and the temperature does not vanish.
Even in this latter case, however, BBs are overwhelmingly 
outnumbered by OOs as we momentarily see. 
Since the heat bath in this case is determined by $K$, not $\Lambda$, we update Equation~(9) 
$\alpha_{BB}=\frac{2\pi m_{0} c^{2}}{\hbar H_{0}\sqrt{-\Omega_{K}}}\gtrsim 1.6\times 10^{69}$ 
where we adopted $\Omega_{K}=-0.058$~\cite{Planck2018} as the lower limit in the case $\Omega_{K}<0$. As pointed out 
below Equation~(9), the specific value $\alpha_{BB}$ is of no relevance for the current 
discussion insofar $\alpha_{BB}\gg 1$.

By virtue of Equation~(2), 
$H_{0}dt=(\Omega_{r}a^{-2}+\Omega_{m}a^{-1}+\Omega_{k}+\Omega_{\Lambda}a^{2})^{-\frac{1}{2}}da$, and correspondingly $d\eta=dt/a$.
In the comoving frame, the volume integrations of \mbox{Equations~(7) and (8)} transform 
as follows $a^{3}(t)d^{3}x dt\rightarrow d^{3}x d\eta$, but 
during DE domination $a\propto e^{Ht}$, where $H$ is constant, 
so $d\eta=da/a^{2}$. On the other hand, $V^{(3)}_{br}\rightarrow V^{(3)}_{br}/a^{3}$, because the brain volume contracts, as is summarized by Equation~(5). 
So, the analog of Equation~(7) in the comoving frame is
\begin{eqnarray}
N_{BB}\sim \frac{V^{(4)}}{V_{br}^{(4)}}\int_{a_*}^{\infty} e^{-a\alpha_{BB}} a da \lesssim \frac{V^{(4)}}{V_{br}^{(4)}}\times\frac{\exp(-\alpha_{BB})}{\alpha_{BB}}\sim\exp(-\alpha_{BB})
\end{eqnarray}
where $\frac{V^{(4)}}{V_{br}^{(4)}}$ is the comoving spacetime volume 
$\int d^{4}x$ (calculated in the relevant region) in units of the four-dimensional brain spacetime volume 
$V_{br}^{(4)}\sim V_{br}^{(3)}\tau$. 
Use has been made in Equation~(12) of the fact that a stupendously large $\alpha_{BB}\gg 1$ is exponentiated, $a_*\gtrsim 1$, and crucially that masses now vary with the scale function, $m_{0}\rightarrow m_{0} a(\eta)$. 
As we see, not even a single BB is expected to form in the comoving frame description during the 
entire eternal history of $\Lambda$CDM, irrespective of the sign of $K$. 
According to our results, $N_{BB}\ll 10^{-69}$ at best. Crucially, it is a combination of the fact that $\alpha_{BB}\gg 1$ and the fact that $m_{0}\rightarrow m_{0}a$ that resulted in $N_{BB}\ll 1$. 

The FOs' case is treated similarly, and as we see it, it is analogously reasoned that 
their abundance is overwhelmed by that of OOs. In this empty Universe case, 
$a=\sqrt{-K}t$, where $K<0$. In this case, $d\eta=da/(\sqrt{-K}a)$ and following 
similar steps to the treatment of the BB case, we obtain 
\begin{eqnarray}
N_{FO}\sim \frac{V^{(4)}}{V_{br}^{(4)}} \int_{a_*}^{\infty} e^{-a\alpha_{FO}} a^2 da \lesssim  \frac{V^{(4)}}{V_{br}^{(4)}}\times\frac{\exp(-\alpha_{FO})}{\alpha_{FO}}\sim\exp(-\alpha_{FO}),
\end{eqnarray}
where $\alpha_{FO}\equiv m_{0}c^{2}\tau/\hbar\gg 1$ and $\tau$ is the 
lifetime of the transient FO. 

Even if we consider a comoving 4-volume as large as the entire observable Universe, 
$\sim$10$^{27}$ cm wide, and an age of 14 Gyrs, and a brain $10$ cm on a side that survives for a mere $\tau\sim 0.1$ s, we obtain the prefactor $\frac{V^{(4)}}{V_{br}^{(4)}}\sim 10^{96}$. It then follows that not even a single BB or FO is expected to form in our four-dimensional comoving Hubble scale volume.  
We conclude that
\begin{eqnarray}
N_{BB}\ll N_{FO}\ll 1\ll 10^{11}<N_{OO}, 
\end{eqnarray}
and the BB and FO paradoxes clearly go away in the comoving frame presentation. 
In the last inequality, use has been made again in the estimate that the cumulative 
human population over history on Earth alone is at least $\sim$120 billion OOs.
In this analysis, we compared the number of observers of each type. Had we 
chosen to compare the number of observations instead, and assuming that OOs 
are much longer lived than BBs or FOs, then a similar (and stronger) inequality 
to Equation~(14) would be obtained. 
To illustrate the dramatic change incurred by analyzing the problem in the comoving frame, it should be noted that even a single electron---let alone a 1 kg brain---is not likely to spontaneously form within 
a volume as large as the comoving Hubble volume over the indefinite 
age of asymptotically de Sitter or Milne universes 
[see Equations (12) and (13)].

The BB and FO paradoxes arise in the cosmic frame essentially 
since the diverging time-integrated three-dimensional comoving volumes in future-eternal 
de Sitter and Milne spacetimes decouple from the corresponding Boltzmann factors. In contrast, in the comoving frame 
picture, masses increase over time, thereby coupling between the Boltzmann factor and the time-integrated three-dimensional comoving volumes. Specifically, whereas BBs contract in this frame, thereby effectively implying three-dimensional volume integration over conformal time, late-time contributions 
are exponentially suppressed; the growing mass BBs and FOs impose natural cutoffs on 
the time integration that is otherwise not imposed in the cosmic frame description. This is so because the 
arguments in the exponents in Equations~(12) and (13) explicitly depend on the scale factor $a$. 

\section{Summary}\label{sec5}

The hypothetical cumulative overabundance of 
Boltzmann Brains (deluded 
disembodied transient brains thermally fluctuating into and out of 
existence due to the asymptotic de Sitter heat bath)
and Freak Observers (cognitively impaired brains that are expected 
to spontaneously form and vanish due to the Heisenberg Uncertainty 
Principle) 
in the $\Lambda$CDM model is a classic example of typicality puzzles 
in cosmology. In a Universe 
where Ordinary Observers are vastly outnumbered by Boltzmann 
Brains or Freak Observers, it becomes exceedingly more likely that we ourselves are 
of either type---entities that are only deluded into believing that they are Ordinary 
Observers. Worse still, observations and theoretical predictions 
made by these cognitively impaired hypothetical observers are 
untrustworthy and undermine the foundations on which their theories rely. 
This unsettling scenario is typically dismissed on the basis of 
unacceptable Cognitive Instability. They are viewed as indicators of 
pathological models, which are rejected primarily due to these troubling 
implications.

The problem with the Cognitive Instability and Atypicality arguments 
is that applying the same standards to the concordance $\Lambda$CDM 
cosmological model leads to its outright disqualification as a 
trustworthy and viable description of our Universe. Yet, it is an undeniably 
well-established fact that $\Lambda$CDM has been an impressively good 
and parsimonious fit to a broad spectrum of independent cosmological 
observations and probes. 

Nevertheless, in an indefinitely expanding Universe---asymptotically 
de Sitter in the case that dark energy is indeed accounted for by a cosmological 
constant or Milne Universe in the case that dark energy eventually decays---the 
cumulative number of Boltzmann Brains and Freak Observers formed after dark energy 
takes over the expansion or alternatively its demise---diverges. In contrast, 
Ordinary Observers have only a finite narrow window in time to 
form in the conventional way (via primordial gas fragmentation, followed in succession by 
galaxy, star, and habitable planetary formation) prior to this epoch, and so their cumulative 
number is finite. The paradox then is how is it that we happen to be of the second 
type whereas our likelihood for being Boltzmann Brains and Freak Observers is exceedingly---actually infinitely---larger. 

As discussed in the present work, this diagnostic tool---together with 
the fact that the concordance $\Lambda$CDM model, described in the 
cosmic frame, implies that we must be atypical observers---perhaps somewhat 
unexpectedly suggests that the Universe is best described as spatially 
static at the background level, while contracting within gravitationally 
bound objects. All this applies, with monotonically evolving masses.

While the cosmic and comoving frame formulations of the 
cosmological model are two observationally 
indistinguishable descriptions of the homogeneous and isotropic 
cosmological model, the finite conformal lifetime of the 
Universe---provided that the latter is indeed described 
by $\Lambda$CDM---is a tantalizing property ideally suited 
to address the Cosmological Constant Problem. This would seem 
in principle to at least suggest a pathway towards 
addressing the Boltzmann Brains and Freak Observers paradoxes 
as they typically form after dark energy comes to 
dominate the expansion dynamics.

However, as was argued here---independent of the Cosmological 
Constant Problem---the abundance of Boltzmann Brains 
and Freak Observers is exponentially suppressed in comparison 
to Ordinary Observers in the comoving frame description. 
This trivial fact appears to have been overlooked in past 
considerations of this foundational and conceptually paradoxical 
aspect  of the standard cosmological model.
This by itself tilts the balance in favor of the comoving frame 
picture over the cosmic frame description; it appears that the 
cognitively stable picture is that space is globally static, 
local gravitationally bound objects monotonically contract, and 
masses monotonically increase. Whereas this dual picture is 
(perhaps for historical reasons) less popular than the expanding 
space perspective, it at least seems to pass---unlike when described in the expanding space picture---the typicality and Cognitive Stability tests. 

\vspace{6pt}

\funding{This research was supported by a grant from the Joan and Irwin Jacobs donor-advised fund at the JCF (San Diego, CA, USA).}

\dataavailability{The original contributions presented in the study are included in the article. Further inquiries can be directed to the authors.}

\acknowledgments{The anonymous referees are gratefully acknowledged for their reviews.}

\conflictsofinterest{The author declares no conflicts of interest. The funders had no role in the design of the study; in the collection, analyses, or interpretation of the data; in the writing of the manuscript; or in the decision to publish the results.}

\newpage
\begin{adjustwidth}{-\extralength}{0cm}
\reftitle{References}

\PublishersNote{}
\end{adjustwidth}

\end{document}